\documentclass[conference]{IEEEtran}
\IEEEoverridecommandlockouts

\usepackage{orcidlink}

\usepackage{hyperref}
\usepackage{amsmath,amssymb,amsfonts}
\usepackage{amsthm}
\usepackage{graphicx}
\usepackage{textcomp}
\usepackage{xcolor}
\newtheorem{definition}{Definition}
\usepackage{todonotes}
\usepackage{comment}

\usepackage[ruled,linesnumbered]{algorithm2e}
\usepackage{algorithmicx}

\usepackage{multirow}
\usepackage{booktabs}
\usepackage{tabularx}
\usepackage{enumitem}
\usepackage{caption}
\usepackage{verbatim}

\usepackage{amssymb}

\usepackage{listings}
\usepackage{xcolor}
\usepackage{diagbox}

\newtheorem{theorem}{Theorem}[section]

\newtheorem{example}[theorem]{Example}

\usepackage{makecell}
\usepackage{balance}

\usepackage{fancyvrb}

\newcommand{\coolname}{$\textsc{FixDrive}$\xspace}

\usepackage{tcolorbox}
\newtcolorbox{promptbox}[1][]{colback=white,colframe=black,sharp corners,#1}

\usepackage[frozencache,cachedir=.]{minted}

\usepackage{tikz}
\usetikzlibrary{shapes,arrows}

\def\BibTeX{{\rm B\kern-.05em{\sc i\kern-.025em b}\kern-.08em
    T\kern-.1667em\lower.7ex\hbox{E}\kern-.125emX}}

\begin{document}

\title{\textsc{FixDrive}: Automatically Repairing Autonomous Vehicle Driving Behaviour for \$0.08 per Violation}

\author{
\IEEEauthorblockN{
    Yang Sun\orcidlink{0000-0002-2409-2160}\textsuperscript{1}\hspace{20pt}
    Christopher M. Poskitt\orcidlink{0000-0002-9376-2471}\textsuperscript{1}\hspace{20pt}
    Kun Wang\orcidlink{0000-0001-5523-1330}\textsuperscript{2}\hspace{20pt}
    Jun Sun\orcidlink{0000-0002-3545-1392}\textsuperscript{1}
}
\IEEEauthorblockA{
    \textsuperscript{1}\textit{School of Computing and Information Systems}, \textit{Singapore Management University}, Singapore\\
    yangsun.2020@phdcs.smu.edu.sg, cposkitt@smu.edu.sg, junsun@smu.edu.sg
}
\IEEEauthorblockA{
    \textsuperscript{2}\textit{State Key Laboratory of Industrial Control Technology}, \textit{Zhejiang University}, Hangzhou, China\\
    kunwang\_yml@zju.edu.cn
}
}

\maketitle

\begin{abstract}
Autonomous Vehicles (AVs) are advancing rapidly, with Level-4 AVs already operating in real-world conditions.
Current AVs, however, still lag behind human drivers in adaptability and performance, often exhibiting overly conservative behaviours and occasionally violating traffic laws.
Existing solutions, such as runtime enforcement, mitigate this by automatically repairing the AV's planned trajectory at runtime, but such approaches lack transparency and should be a measure of last resort.
It would be preferable for AV repairs to generalise beyond specific incidents and to be interpretable for users.
In this work, we propose \coolname, a framework that analyses driving records from near-misses or law violations to generate AV driving strategy repairs that reduce the chance of such incidents occurring again.
These repairs are captured in $\mu$Drive, a high-level domain-specific language for specifying driving behaviours in response to event-based triggers.
Implemented for the state-of-the-art autonomous driving system Apollo, \coolname identifies and visualises critical moments from driving records, then uses a Multimodal Large Language Model~(MLLM) with zero-shot learning to generate $\mu$Drive programs.
We tested \coolname on various benchmark scenarios, and found that the generated repairs improved the AV's performance with respect to following traffic laws, avoiding collisions, and successfully reaching destinations.
Furthermore, the direct costs of repairing an AV---15 minutes of offline analysis and \$0.08 per violation---are reasonable in practice.

\end{abstract}

\section{Introduction}
Autonomous Vehicles (AVs) are currently undergoing rapid and promising development. Notably, several Level-4 AVs, which do not require driver intervention, have been successfully deployed in real-world traffic scenarios~\cite{sae2018taxonomy}. Prominent examples include Google Waymo~\cite{waymo}, Baidu Apollo~\cite{apolloauto}, and TuSimple~\cite{TuSimple}. These AVs are capable of performing critical tasks such as perception, trajectory planning, and actuation control. However, despite these advancements, AVs are far from perfect and still lag significantly behind human drivers in terms of performance and adaptability. 
For instance, AVs sometimes exhibit overly conservative driving behaviour, which can lead to situations where they become stuck on the road~\cite{wan2022too}. 
Furthermore, Autonomous Driving Systems~(ADSs)---the `brains' of AVs, responsible for perception, decision-making, and control---can also be overly aggressive and cause accidents under specific conditions~\cite{Bashetty20DeepCrashTest, li2020av, Sun-Poskitt-et_al22a, Zhou-et_al23a}.  Such behaviours are often easily recognisable and avoidable by human drivers, underscoring the need for significant improvements before AVs can match or surpass human driving capabilities.

Existing work offers two categories of solutions to address these problems. The first category uses rule-based runtime enforcement to correct problematic behaviour directly~\cite{sun2024redriver, Grieser-et_al20a, hong2020avguardian, Cheng-et_al21a, Shankar-et_al20a, Mauritz-et_al16a, d2005lola, Watanabe-et_al18a}. For example, when an ADS encounters potential violations of specific property specifications (e.g.~traffic laws), one proposed solution, outlined in REDriver~\cite{sun2024redriver}, uses a gradient-based algorithm to modify the AV's planned trajectory. 
However, these repairs are limited in scope and lack transparency, since they are very low-level and difficult for users to interpret.
Furthermore, they are meant to be a measure of last resort rather than a general correction to driving strategies.

The second category involves learning-based methods, which train ADSs to behave like human drivers using real driving data. These approaches focus on exploring and summarising human driver behaviour patterns to guide the driving modes of ADSs~\cite{chen2023end, prakash2021multi}, such as imitation learning to replicate expert behaviour and train the ADS to drive in a human-like manner~\cite{sama2020extracting, wei2010learning, xu2015establishing, xu2020learning, le2022survey}. However, these approaches often fall short due to the difficulty of capturing the nuanced decision-making processes of human drivers from limited data, leading to poor generalisation or overfitting to specific tasks. Consequently, there is a need for an AV driving strategy repair approach that generalises beyond specific incidents and is interpretable for users.

Multimodal Large Language Models (MLLMs) appear to be intelligent and ideally suited for improving ADSs due to their advanced text and image understanding and reasoning capabilities~\cite{brown2020language, achiam2023gpt, yin2023survey}. Trained on massive datasets, MLLMs can interpret and replicate human driving behaviour, thereby making ADS decisions more explainable~\cite{cui2024survey}.
Existing works (e.g.~\cite{chen2023driving, mao2023gpt, wen2024road}) explore utilising MLLMs to replace parts of the ADS, such as perception, planning, and control, thereby making the decision-making logic more understandable. For example, GPT-Driver~\cite{mao2023gpt} abstracts the perception and prediction results of the ADS into language tokens, then uses OpenAI GPT 3.5 to directly produce the planned trajectory along with explanations.
However, the inherent latencies and uncertainty associated with generative models make it impractical to build an ADS based solely on online MLLMs.
Additionally, there is a significant gap between natural language and the actual control commands for autonomous vehicles, making it challenging to directly apply MLLM-generated solutions to real-world driving tasks.
Currently, there are no practical approaches that leverage MLLMs in a manner that is both offline and compatible with existing ADS frameworks such as Apollo~\cite{apolloauto} and Autoware~\cite{autoware}.

\begin{figure}
    \centering
    \includegraphics[width=\linewidth, trim={5cm 6cm 5cm 5cm}, clip]{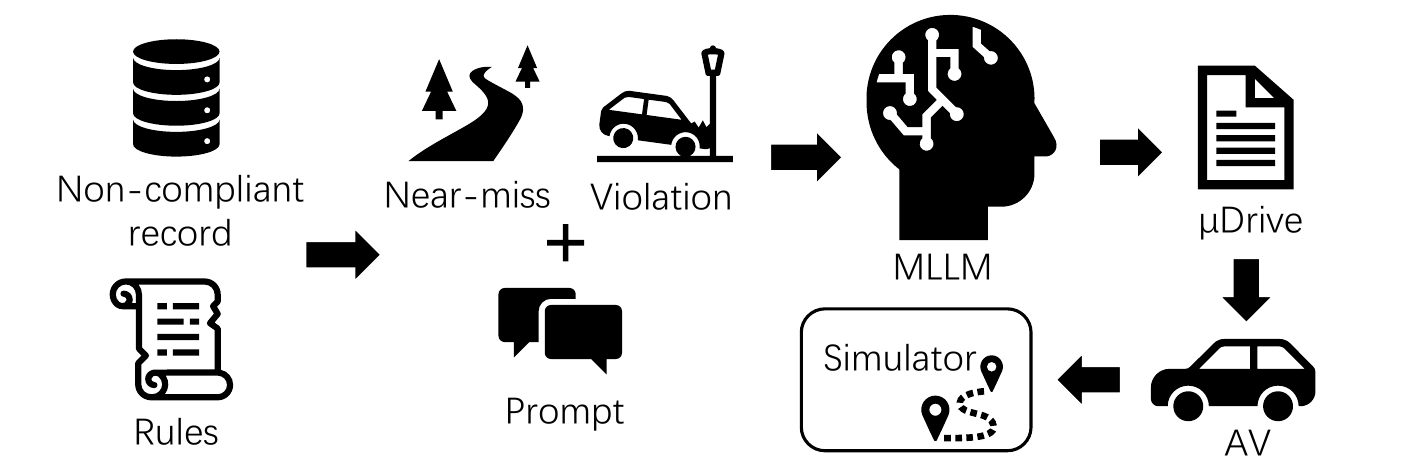}
    \vspace{-0.4cm}
    \caption{Overview of \coolname}
    \label{fig:overflow}
    \vspace{-0.6cm}
\end{figure}

In this work, we propose \coolname, a method that analyses records (i.e.~comprehensive log files) from bad driving behaviours such as collisions, near misses, or law violations, then generates general AV driving strategy repairs to reduce the chance of such incidents occurring again.
Rather than modifying code~\cite{LeGouesDFW12} or applying opaque low-level fixes~\cite{sun2024redriver}, \coolname produces repairs in $\mu$Drive~\cite{wang2024mudrive}, a high-level Domain-Specific Language~(DSL) for specifying the driving behaviours that should occur upon certain triggers (e.g.~approaching a traffic light).
\coolname identifies and visualises critical moments from incident records, then utilises an MLLM with zero-shot learning to generate $\mu$Drive programs that repair the driving strategy.
This translation is executed offline and once per violation, allowing our approach to leverage the reasoning capabilities of MLLMs while mitigating their latency issues.
Additionally, by generating repairs in a high-level DSL, they are more interpretable compared to those of low-level gradient-based approaches like REDriver~\cite{sun2024redriver}.
Note that we categorise methods as either offline or online based on how they interact with runtime driving decisions. For example, an MLLM that generates real-time driving decisions based on the current driving context is classified as an online method. Conversely, \coolname is an offline framework, enhancing the ADS through repair scripts generated \emph{before} further runs.

An overview of \coolname is shown in Figure~\ref{fig:overflow}.
Users need only provide records from executed driving scenarios and the corresponding property specifications (e.g.~traffic laws, collision avoidance) that were violated.
\coolname automatically identifies two critical moments from the records (the `near miss' and the `violation' moments), visualises them for a multimodal prompt, then utilises a state-of-the-art MLLM (OpenAI GPT 4~\cite{openaichatgpt}) to generate driving strategy repair scripts in the $\mu$Drive~\cite{wang2024mudrive} DSL.
Additionally, OpenAI's function calling~\cite{gpt_function_calling} is used to ensure that the MLLM generates syntactically valid $\mu$Drive programs, which are specified via a JSON Schema.
The resulting program is then applied to the Apollo ADS~\cite{apollo90}, dynamically adjusting the parameter settings of the planning module to repair its driving strategy at runtime.

We evaluated \coolname on a set of benchmark scenarios in which the ADS violated various property specifications, such as different traffic laws, collision avoidance, and successful journey completion. \coolname provided effective general driving strategy repairs that helped the ADS successfully navigate these problematic scenarios without adversely affecting performance in normal scenarios. Additionally, \coolname consistently generated effective AV driving strategy repairs with practically reasonable direct costs, i.e.~less than 15 minutes (and typically around 10 minutes) of offline analysis and \$0.08 per violation.

\section{Background and Problem}
\label{sec:overview_and_background}
In this section, we review the architecture of ADSs, DSLs for specifying safety properties and specifying high-level driving behaviours, the current capabilities of MLLMs, and then formally define our problem.

\subsection{Overview of ADSs}
State-of-the-art open-source ADSs such as Apollo~\cite{apollo90} and Autoware~\cite{autoware} share similar architectures. These systems are typically organised into loosely-coupled modules that communicate via message-passing. Three of these modules are particularly relevant in our context: perception, motion planning, and control.

The perception module receives sensor readings (e.g.~from cameras or LiDAR), processes them, and publishes the resulting data to the motion planning module. The motion planning module then classifies the current driving scenario into categories such as \emph{lane follow}, \emph{borrow lane}, and \emph{traffic light handling}. Each scenario has distinct processing logic and key parameters. For example, the \emph{emergency pull-over} scenario involves two key parameters: \emph{Expected\_speed} and \emph{Stopping\_distance}. During an emergency pull over, the vehicle is expected to rapidly decrease to the \emph{Expected\_speed} and then proceed to pull over, with the \emph{Stopping\_distance} indicating how far the vehicle should travel before coming to a complete stop. For more detailed information on how these key parameters work, we refer the reader to~\cite{apollo90, autoware}.

For each scenario, the motion planning module generates a corresponding \emph{planned trajectory} based on the map, destination, sensor inputs, and the state of the \emph{ego vehicle} (i.e.~the vehicle under ADS control). This planned trajectory outlines the vehicle's future positions at various time points, taking into account the predicted environment, which includes factors such as the anticipated movements of other vehicles (NPCs), pedestrians, and traffic light states.
Finally, the control module translates the planned trajectory into control commands (e.g.~braking, acceleration, steering, and signaling) so that the ego vehicle follows the planned trajectory, passing through the waypoints with the desired speed, acceleration, steering angle, and gear position. 

\begin{figure}[t]
    \centering\footnotesize
    \begin{align*}
    \phi\;:=&\;\mu \;
     |\;\neg\phi\;
     |\;\phi_1 \lor \phi_2\;|\; \phi_1 \land \phi_2\;|\; \phi_1 \;\mathtt{U}_I\; \phi_2\;\\
    \mu\;:=&\;f(x_0,x_1,\cdots, x_k) \sim 0 \ \ \sim \;:= > \;| \geq\;| <\;| \leq\;| \neq\;| =;
    \end{align*}
    \captionsetup{skip=0pt}
    \caption{Specification language syntax, where $\phi$, $\phi_1$ and $\phi_2$ are STL formulas, $I$ is an interval, and $f$ is a multivariate linear continuous function over language variables $x_i$}
    \label{syntax}\vspace{-10pt}
\end{figure}

\subsection{Specifying Safety Properties}
\label{sec:Specification definition}
In the context of AVs, safety should not simply mean the absence of collisions, but also adherence to the rules of the road that drivers are supposed to abide by.
To that end, we adopt the property specification language used by LawBreaker~\cite{Sun-Poskitt-et_al22a}, as well as the project's existing specifications of the traffic laws of China and Singapore.
The specification language is based on Signal Temporal Logic~(STL), and is evaluated with respect to traces of scenes, providing a way to automatically determine whether a tester-defined property was violated or not in a simulated run of the ADS.
We highlight the key features of the specification language below (the full syntax and semantics is given in~\cite{Sun-Poskitt-et_al22a}).

The high-level syntax of the language is shown in Figure~\ref{syntax}.
A time interval $I$ is of the form $[l,u]$, where $l$ and $u$ are respectively the lower and upper bounds of the interval. 
Following convention, we write $\Diamond_I \; \phi$ to denote $true \; \mathtt{U}_I \; \phi$; and $\Box_I \; \phi$ to denote $\neg \; \Diamond_I \; \neg \phi$. Intuitively, $\mathtt{U}$, $\Box$, and $\Diamond$ are modal operators that are respectively interpreted as `until', `always', and `eventually'.
We omit the time interval when it is $[0, \infty]$.

In general, $\mu$ can be regarded as a proposition of the form $f(x_0,x_1,   \cdots, x_k) \sim 0$, where $f$ is a multivariate function and $x_i$ for all $i$ in $[0,k]$ is a variable supported in the language.
\begin{example}
    Suppose we have a signal variable $speed = \langle speed(0),   speed(1), \dots, speed(n) \rangle$, which represents the autonomous vehicle's speed throughout its journey.
    Then, we can create a simple Boolean Expression $\mu = speed(t) < 60$ to test whether the speed of the vehicle is larger than $60km/h$. 
    Note that $\mu$ can be regarded as a proposition of the form $60 - speed(t) > 0$ or $speed(t) - 60 < 0$.
    To verify whether $\mu$ holds true at all time steps, we can use the temporal logic symbol `always', resulting in the formula $\varphi = \Box (speed < 60)$.
\end{example}

A specification is evaluated with respect to a trace $\pi$ of \emph{scenes}, denoted as $\pi=\langle \pi_0, \pi_1, \pi_2 \ldots, \pi_n \rangle$, where each scene $\pi_i$ is a valuation of the propositions at time step $i$, and $\pi_0$ reflects the state at the start of a simulation.
The language follows the standard semantics of STL (see e.g.~\cite{maler2004monitoring}).

\begin{figure}[t]
\begin{center}
\renewcommand{\arraystretch}{1.2}\footnotesize
\begin{tabular}{lcl}
program & ::= & \{rule\}+~ \\
rule & ::= &  '$\mathtt{rule}$' string\_literal \\
  && '$\mathtt{trigger}$' event\_trigger \\
  && ['$\mathtt{condition}$' \{['$\mathtt{!}$'] condition\}+] \\
  && '$\mathtt{then}$' \{action\}+~ \\
  && ['$\mathtt{until}$' event\_trigger] \\
  && '$\mathtt{end}$' \\
event\_trigger & ::= & event $\mid$ '$\mathtt{always}$' \\
\end{tabular}
\end{center}
\vspace{-0.2cm}
	\caption{Abstract syntax of $\mu$Drive programs}
	\label{fig:abstract_syntax}
 \vspace{-0.5cm}
\end{figure}

\subsection{Specifying Driving Behaviours}
\label{sec:udrive grammar}
The default output of a large language model (LLM) is natural language, which can be vague and challenging to utilise. To obtain specific and actionable behaviours for AVs, we need a robust method to ensure that the output of the LLM is always valid and directly applicable to AVs.
To achieve this, we utilise the high-level DSL $\mu$Drive~\cite{wang2024mudrive}, which allows driving behaviours to be specified in simple rules that are triggered by contextual events (e.g.~approaching a traffic light). 

The abstract syntax of $\mu$Drive in EBNF format is shown in Figure~\ref{fig:abstract_syntax}.
A $\mu$Drive program contains one or more rules, each consisting of up to five parts: 
1) a \emph{name} or description expressed as a string;
2) a \emph{trigger}, which is an event that causes the rule to be applied;
3) zero or more \emph{conditions}, which constrain the application of the rule;
4) one or more \emph{actions}, which are assignments of driving-related variables that are applied for the duration of the rule;
5) at most one \emph{exit trigger}, which is an event that ends the application of the rule.

\begin{figure}[t]
\begin{minted}[fontsize=\scriptsize,frame=single,obeytabs=true,tabsize=4,numbersep=-10pt,escapeinside=||]{yaml}
|\textbf{\textcolor{green!50!black}{rule}}| |\textcolor{red!70!black}{"Drive slowly through a junction when there is }
\textcolor{red!70!black}{an obstacle."}|
|\textbf{\textcolor{green!50!black}{trigger}}| 
    entering_junction 
|\textbf{\textcolor{green!50!black}{condition}}| 
    obstacle_distance_leq(20)
    is_traffic_light(green)
|\textbf{\textcolor{green!50!black}{then}}|
    cruise_speed(30)
|\textbf{\textcolor{green!50!black}{until}}|
    exiting_junction
|\textbf{\textcolor{green!50!black}{end}}|
    \end{minted}
  \caption{\textsc{$\mu$Drive} driving strategy repair example}
  \label{lst:example}
  \vspace{-0.5cm}
\end{figure}

Intuitively, events represent states monitored by $\mu$Drive as the AV drives through its environment. For example, the events $\mathtt{entering\_junction}$ and $\mathtt{exiting\_junction}$ are set to \texttt{True} when the AV is entering or exiting a junction, respectively.
Conditions specify what must be true of the current environment to allow the rule to be applied. For example, the conditions $\mathtt{is\_traffic\_light(green)}$ and $\mathtt{obstacle\_distance\_leq(20)}$ indicate that the rule can take effect only when the traffic light ahead is green and the AV is within 20 metres of an obstacle.
Actions are tasks executed throughout the duration of a rule application. For example, the action $\mathtt{cruise\_speed(30)}$ sets the default planning speed of the AV to 30 km/h.
An overall example of a $\mu$Drive program is shown in Figure~\ref{lst:example}. This program indicates that within a junction, if an obstacle is detected within 20 metres and the traffic light ahead is green, the default planning speed of the AV should be set to 30 km/h.
For a detailed introduction to the grammar of $\mu$Drive, we refer readers to \cite{wang2024mudrive}.

\subsection{Multimodal Large Language Models (MLLMs)}
MLLMs integrate and process multiple types of data, including text, images, audio, and video. These models utilise the capabilities of large-scale neural networks to comprehend and generate content across different modalities, thereby offering more comprehensive and versatile AI functionalities.
State-of-the-art MLLMs, such as OpenAI's GPT-4~\cite{openaichatgpt} and Google's PaLM-E~\cite{driess2023palm}, exemplify the advancements in this field. GPT-4, for instance, can process textual inputs while simultaneously understanding and interpreting images, enabling it to describe images, answer related questions, and seamlessly integrate visual information with textual content. Similarly, PaLM-E, a large multimodal embodied language model, integrates textual and visual data to enhance its ability to comprehend and interact with the physical environment.

These models are trained on extensive datasets that encompass diverse forms of media, which allows them to acquire a vast amount of general knowledge. This training enables MLLMs to perform a wide variety of tasks, such as generating detailed image captions, providing contextually aware responses, and enhancing search engines with improved understanding of visual queries.
The multimodal approach significantly enhances the model's utility, making it capable of tasks beyond the scope of single-modality models. By integrating multiple types of data, MLLMs are advancing the frontier of AI, creating more intuitive and intelligent systems that better mimic human cognition and understanding.

\subsection{The Problem}
Assuming the availability of a powerful MLLM, such as ChatGPT, with the capability to comprehend driving conditions and provide appropriate suggestions, the challenge lies in efficiently leveraging this MLLM to analyse records of AV violations and generate driving strategy repairs in $\mu$Drive that would prevent such violations occurring again in the future.
We formulate our problem as follows:

\begin{definition}[Problem Definition]
\label{def:problem_def}
    Given an MLLM, a record $\alpha$ of the AV in a scenario, and a property specification of ADS behaviour $\varphi$, suppose that $\alpha \nvDash \varphi$.
    The objective is to utilise the MLLM to generate $\mu$Drive programs based on the combination of $\alpha$ and $\varphi$. 
    Let $\alpha'$ denote the resulting record of the AV by replaying the same scenario with the $\mu$Drive programs generated by the MLLM applied.
    The goal is to increase the likelihood of $\alpha' \vDash \varphi$.
\end{definition}

Intuitively, when a scenario is identified in which the AV violates specified properties, we provide relevant information in the prompt to the MLLM to enable it to understand the current situation. The MLLM then offers suggestions in the form of a $\mu$Drive program to help ensure such a violation does not occur again. To maintain minimal and interpretable additional control logic, the $\mu$Drive programs should be kept as small as possible.
This context highlights two critical requirements for our approach:
1) develop a method to automatically provide accurate and relevant information to the MLLM;
2) ensure that the MLLM's suggestions are translated into valid and effective $\mu$Drive programs.

\section{Our Approach}
\label{sec:our_approach}
\coolname, our framework for obtaining general driving strategy repairs via MLLMs, comprises three main steps. First, \emph{problem localisation}, which identifies when the violation of the specified properties occurred and any near-miss situations. Second, \emph{prompt generation}, which automatically generates the necessary text prompts and visualisations of specific driving conditions. These specific conditions refer to the time steps when property violations and near-miss situations occurred. Finally, \emph{$\mu$Drive script generation}, which formats the MLLM's responses into syntactically valid $\mu$Drive scripts, ensuring the driving strategy repairs are executed by the ADS.

\subsection{Problem Localisation}
One possible way to allow a language model to comprehend a driving incident is to provide it with a complete record, i.e., a structured log file that captures all necessary data to recreate the driving scenario. This log would include detailed information on the driving environment perception, routing details, predictions of other vehicles, pedestrians, and traffic lights, as well as motion planning and control commands.
However, processing such comprehensive records is computationally intensive, costly, and time-consuming. Fortunately, in driving scenarios where certain properties are violated, there are always a few key moments that hold significant importance. Identifying these moments allows for a better understanding of the driving scenario, and can also be automated.
For example, ACAV~\cite{sun2024acav} reduces the length of driving records by 62.23\% based on a causality analysis, and correctly identifies causal events in 93.64\% of a set of generated accident records.

In this work, we develop a lightweight approach---based on a quantitative semantics---to identify two critical moments: the \emph{near-miss} and \emph{violation moments}. The violation moment is the point at which a specific property is violated. The near-miss moment occurs a few time steps before the violation, during which the violation is likely but has not yet happened. For example, if the property that the AV should follow is `avoid collisions with other objects', the violation moment is when the vehicle collides with another object, while the near-miss moment is when the vehicle fails to maintain a safe distance from other vehicles. The logic is straightforward: the violation moment represents the final outcome, whereas the near-miss moment could be the potential cause of the property violation.
We explain in the following how to identify them.

\noindent \emph{\textbf{Quantitative Semantics.}}
\label{sec:Quantitative_Semantics}
To locate critical moments, we need a method for quantitatively evaluating whether the current trace satisfies a given property. To achieve this, given a property specification $\varphi$ and a driving record for the ADS, \coolname first constructs a trace $\pi$ from the record by evaluating all variables relevant to $\varphi$ at each time point. 
An execution trace $\pi$ is a sequence of scenes, denoted as {\small $\pi=\langle \theta_0, \theta_1, \ldots, \theta_n \rangle$}. A scene $\theta$ is a tuple of the form {\small $\theta=( f_0, f_1, \ldots, f_x )$}, where each $f_i$ is the valuation of a variable.
These variables describe the status of the vehicle, traffic signal states, and traffic conditions. For example, variables such as `isOverTaking', `junctionAhead(n)', and `NPCAhead(n)' indicate whether the vehicle is changing lanes, whether there is a junction ahead within \emph{n} metres, and whether there is a vehicle ahead within \emph{n} metres, respectively.
For a detailed introduction to these variables, we refer the readers to \cite{Sun-Poskitt-et_al22a}.

Next, \coolname computes how `close' the ego vehicle will come to violating $\varphi$.
To measure how close a trace $\pi$ is to violating $\varphi$, we adopt a quantitative semantics~\cite{maler2004monitoring, deshmukh2017robust,nivckovic2020rtamt} that produces a numerical \emph{robustness} degree. 

\begin{definition}[Quantitative Semantics]\label{def:Quantitative_Semantics}
Given a trace $\pi$ and a formula $\varphi$, the quantitative semantics is defined as the robustness degree $\rho(\varphi, \pi,t)$, computed as follows.
Recall that propositions $\mu$ are of the form $f(x_0,x_1,\cdots,x_k) \sim 0$.
{\small\begin{equation*}
  \rho(\mu, \pi, t) =
    \begin{cases}
      -\pi_t(f(x_0,x_1,\cdots, x_k)) & \text{if $\sim$ is $\leq$ or $<$}\\
      \pi_t(f(x_0,x_1,\cdots, x_k)) & \text{if $\sim$ is $\geq$ or $>$}\\
      \mid \pi_t(f(x_0,x_1,\cdots, x_k)) \mid & \text{if $\sim$ is $\neq$}\\
      -\mid \pi_t(f(x_0,x_1,\cdots, x_k)) \mid & \text{if $\sim$ is $=$}
    \end{cases}       
\end{equation*}}

\noindent where $t$ is the time step and $\pi_t(e)$ is the valuation of expression $e$ at time $t$ in $\pi$.
{\small\begin{align*}
\rho(\neg\varphi,\pi,t) & = -\rho(\varphi,\pi,t) \\
\rho(\varphi_1 \land \varphi_2,\pi,t) & = \min\{\rho(\varphi_1,\pi,t),\rho(\varphi_2,\pi,t)\} \\
\rho(\varphi_1 \lor \varphi_2,\pi,t) & = \max\{\rho(\varphi_1,\pi,t),\rho(\varphi_2,\pi,t)\} \\
\rho(\varphi_1 \;\mathtt{U_I}\; \varphi_2,\pi,t) & = \sup_{t_1 \in t+\mathtt{I}} \min \{\rho(\varphi_2,\pi,t_1), \inf_{t_2 \in [t,t_1]} \rho(\varphi_1,\pi,t_2)\}\\
\rho(\Diamond_\mathtt{I}\varphi,\pi,t) & = \sup_{t'\in t+\mathtt{I}}\rho(\varphi,\pi,t') \\
\rho(\Box_\mathtt{I}\varphi,\pi,t) & = \inf_{t'\in t+\mathtt{I}}\rho(\varphi,\pi,t') \\
\rho(\bigcirc \varphi,\pi,t) & = \rho(\varphi,\pi,t+1)
\end{align*}}
where $t+I$ is the interval $[l+t,u+t]$ given $I=[l,u]$.
\qed
\end{definition}

Note that the smaller $\rho(\varphi,\pi,t)$ is, the closer $\pi$ is to violating $\varphi$.
If $\rho(\varphi,\pi,t) \leq 0$, $\varphi$ is violated.
We write $\rho(\varphi,\pi)$ to denote $\rho(\varphi,\pi,0)$; $\pi \vDash \varphi$ to denote $\rho(\varphi,\pi,t) > 0$; and $\pi \not \vDash \varphi$ to denote $\rho(\varphi,\pi,t) \leq 0$. Note that time is discrete in our setting. 

\begin{example}
\label{example:robustness calculation}
Let $\varphi = \Box (speed < 60)$, i.e.~the speed limit is $60$km/h.
Suppose $\pi$ is $\langle (speed \mapsto 0, \dots), (speed \mapsto 0.3, \dots), \cdots  (speed \mapsto 50, \dots) \rangle$ where the ego vehicle's max $speed$ is $50$km/h at the last time step.
We have $\rho(\varphi, \pi) = \rho(\varphi, \pi, 0) = min_{t \in [0, |\pi|]} ( 60 - \pi_t(speed) ) = 10$.
This means that trace $\pi$ satisfies $\varphi$, and the robustness value is 10. 
\qed
\end{example}

\noindent \emph{\textbf{Violation and Near-Miss Moments.}}
\label{sec:Violation and Near-Miss Moment}
With quantitative semantics, we can now introduce the method to locate the \emph{violation moment} and \emph{near-miss moment}.
Given a trace $\pi = \langle \pi_0, \cdots, \pi_n \rangle$, let $\pi^k$ denote the prefix $\langle  \pi_0, \cdots, \pi_k \rangle$, where $k \leq n$. Intuitively, $\pi^k$ represents the first $k$ time steps of the original trace $\pi$.
For the \emph{violation moment}, we identify the smallest $k$ that satisfies $\rho(\varphi, \pi^k) \leq 0$.
For the \emph{near-miss moment}, we adopt a user-customisable threshold $\delta$. We aim to identify a time step $k$ such that $\rho(\varphi, \pi^k) \leq \delta$ and there does not exist a time step $l$ such that $l < k$ and $\rho(\varphi, \pi^l) < \delta$. Note that $\delta$ is determined empirically  in our evaluation (see Section~\ref{sec:implementation_evaluation}).
Intuitively, $k$ is the earliest time step when the robustness value falls below the threshold $\delta$. We identify the time step $k$ using a sequential search, starting from $k=0$ and incrementing $k$ until we find a $k$ such that $\rho(\varphi, \pi^k) < \delta$.

\begin{example}
\label{example:timestep location}
Let $\varphi = \Box (speed < 60)$, i.e.~the speed limit is $60$km/h. Suppose the threshold $\delta = 5$
Suppose $\pi$ is $\langle \pi_0 = (speed \mapsto 0, \dots), \pi_1 = (speed \mapsto 1, \dots), \cdots \pi_{90} = (speed \mapsto 90, \dots) \rangle$ where the ego vehicle $speed$ is increasing over time steps and the ego vehicle's max $speed$ is $90$km/h at the last time step.
We have $\rho(\varphi, \pi) = \rho(\varphi, \pi, 0) = min_{t \in [0, |\pi|]} ( 60 - \pi_t(speed) ) = -30$. Hence, the specification is violated. 
The following are computed in sequence:
\begin{align*}
    & \rho(\varphi, \pi^{0}) = 60, \rho(\varphi, \pi^1)  = 59, \cdots,
     \rho(\varphi, \pi^{55}) = 5, \cdots, \\
    & \rho(\varphi, \pi^{59}) = 1, \rho(\varphi, \pi^{60}) = 0,  
    \rho(\varphi, \pi^{61}) = -1, \cdots
\end{align*}

To identify the violation moment, we find the smallest time step $k$ where $\rho(\varphi, \pi^k) \leq 0$. In this case, $k = 60$. Similarly, the smallest time step $k$ where $\rho(\varphi, \pi^k) \leq 5$ is $k = 55$.
Therefore, the violation moment occurs at time step 60, and the near-miss moment occurs at time step 55.
\qed
\end{example}

\subsection{Prompt Generation}
\tikzstyle{box} = [rectangle, draw, text width=1.8cm, text centered, rounded corners, minimum height=1.5cm]
\tikzstyle{arrow} = [thick,->,>=stealth]

The input for an MLLM can be in various formats, such as images, videos, audio, and text prompts. Given that MLLMs are trained on extensive datasets rich in knowledge, we anticipate they will `understand' the prompts we provide, much like an intelligent human. In this work, we utilise two types of prompts: visualisations of driving conditions and text descriptions to convey essential information not covered by the visualisations.

\begin{figure*}[htbp]
    \centering
    \includegraphics[width=\textwidth]{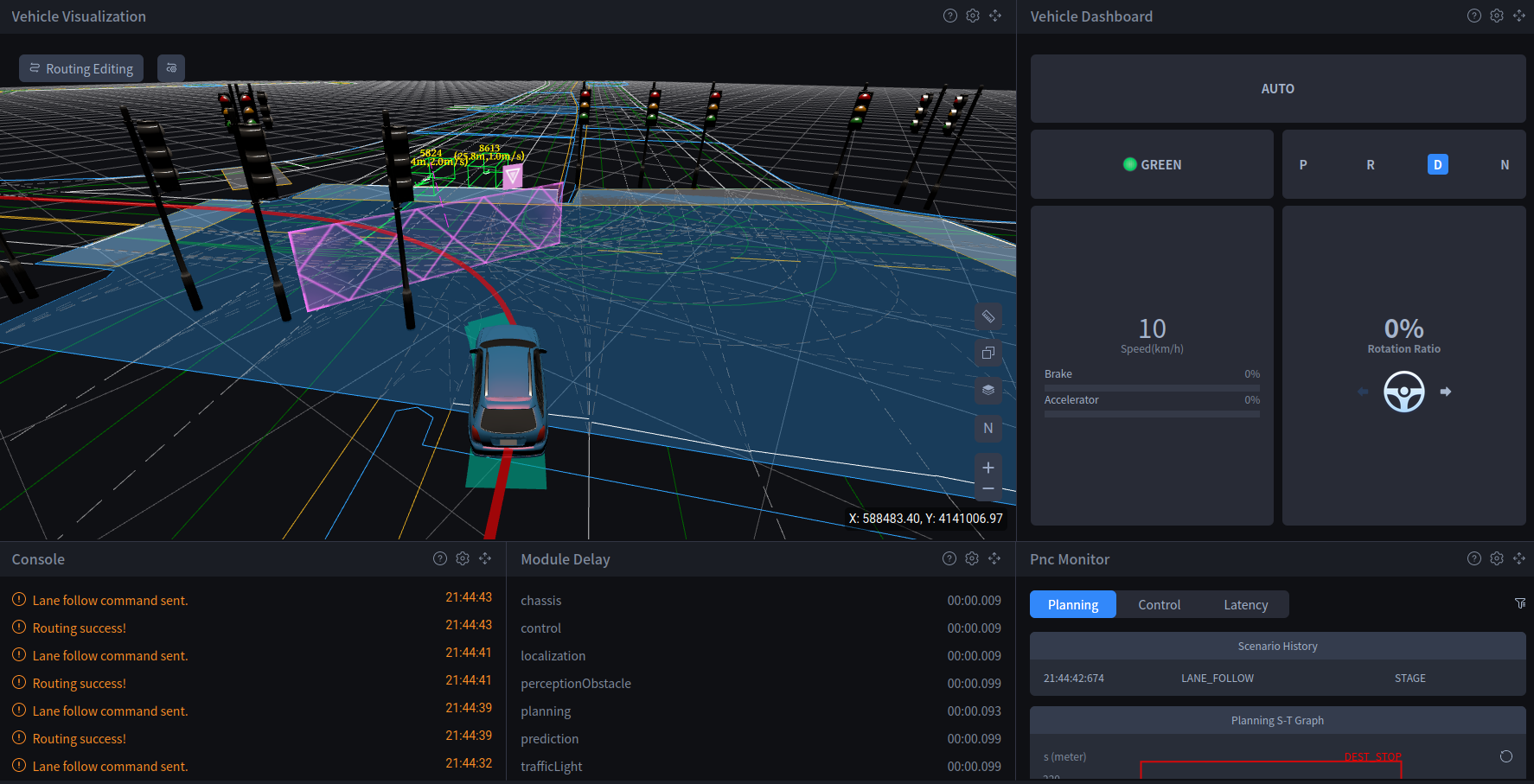}
    \caption{Visualisation of a scenario, which is provided to the MLLM along with an `overall prompt'} 
    \label{fig:visualisation} 
    \vspace{-0.6cm}
\end{figure*}

\noindent\emph{\textbf{Visualisations of Scenarios.}}
In the driving records of ADS trajectories, each time step contains extensive information such as the speed, acceleration, and steering angle of the AV, as well as the positions of other vehicles and pedestrians. While it is possible to describe this information in natural language, it does not provide a direct impression of the driving scenario. For example, given the positions of the AV and another background vehicle, it can be challenging to determine the exact direction of the background vehicle relative to the AV.

Fortunately, visualising the driving scenario can help alleviate this problem, and state-of-the-art ADSs, such as Apollo, offer this capability. Detailed information, including the positions of various objects, can be effectively conveyed through visualisation by displaying a grid map that shows the relative positions of each object.

An example of this visualisation is shown in Figure~\ref{fig:visualisation}. In this visualisation, the upper-left section, labelled `Vehicle Visualization', displays the current driving conditions of the AV (marked in blue), other vehicles (marked in green boxes), pedestrians (marked in yellow boxes), cyclists (marked in blue boxes), and unknown objects (marked in purple boxes). Each box includes numerical values indicating the distance to the AV and the current speed of the object. The predicted trajectory of each object is shown as a coloured line.
The lower-left section, labelled `Console', shows logs from the ADS, while the `Module Delay' section indicates the delay of each module.
The right section, labelled `Vehicle Dashboard', shows the current status of the AV and the detected status of traffic lights ahead. The `Pnc Monitor' section provides detailed information on the inner decisions of the planning and control modules.

This visualisation compactly encodes rich information in a human-friendly manner.
This is important for the transparency of the approach: it allows the MLLM's decisions to be based on the same high-level information that drivers work with, instead of (for example) low-level gradient-based discrepancies.

\noindent \emph{\textbf{Overall Prompt.}}
The overall prompt consists of two parts: the first part includes two images illustrating the visualisation of the \emph{violation moment} and the \emph{near-miss moment} as shown in Section~\ref{sec:Violation and Near-Miss Moment}, and the second part is a text prompt.

Our text prompts complement the ADS visualisation by providing necessary additional information. These prompts follow a specific workflow to enable automatic generation. 
First, we specify an identity for the MLLM:
\vspace{-0.15cm}\begin{promptbox}\small
\vspace{-0.15cm}
(identity) Suppose you are a driver.
\vspace{-0.15cm}
\end{promptbox}\vspace{-0.15cm}

Next, since the visualisations do not contain weather information, we provide information about the current weather conditions. For example, we state the following if there is no rain, fog,  or snow, and the visibility is more than 50 metres:
\vspace{-0.15cm}\begin{promptbox}\small
\vspace{-0.15cm}
(weather) There is nothing noteworthy about the weather.
\vspace{-0.15cm}
\end{promptbox}\vspace{-0.15cm}
To aid the MLLM's understanding of the provided image, we offer background details about the input image. This includes describing different aspects of the visualisation to help the MLLM to understand it:
\vspace{-0.15cm}\begin{promptbox}\small
\vspace{-0.15cm}
(background) In these pictures, the left side shows the visualisation of the driving record. 
The right side displays the status of the traffic light, vehicle speed, and steering angle. 
The green boxes indicate detected vehicles, yellow boxes indicate detected pedestrians, blue boxes indicate detected bicycles, and purple boxes indicate unknown objects.
\vspace{-0.15cm}
\end{promptbox}\vspace{-0.15cm}
Following this, we describe the property specification that the AV is supposed to satisfy, e.g.~the avoidance of collisions or adherence to traffic laws. 
Note that we use the original traffic laws from~\cite{China_traffic_law} as input when the property is an STL-based traffic law specification.
For example, if we are testing the property `no collisions', we would state:

\vspace{-0.15cm}\begin{promptbox}\small
\vspace{-0.15cm}
(rule) You are supposed to follow the following rule: Avoid collisions with other objects.
\vspace{-0.15cm}
\end{promptbox}\vspace{-0.15cm}
\vspace{-0.08cm}
Then, we specify the time gap between the \emph{violation moment} and the \emph{near-miss moment}, and clarify that the former image depicts the violation:
\vspace{-0.15cm}\begin{promptbox}\small
\vspace{-0.15cm}
(sequence) The second picture was taken 4 seconds later than the first picture, capturing the moment when the rule violation occurred.
\vspace{-0.15cm}
\end{promptbox}\vspace{-0.15cm}
Finally, we provide some initial settings of the current ADS to assist the MLLM in making decisions:
\vspace{-0.15cm}\begin{promptbox}\small
\vspace{-0.15cm}
(default) In the original ADS, the initial settings are: $max\ planning\ speed = 72 km/h$, ...
\vspace{-0.15cm}
\end{promptbox}\vspace{-0.15cm}

\subsection{$\mu$Drive Script Generation}
The driving strategy repairs generated by the MLLM must be in the correct format, i.e.~pertaining to the $\mu$Drive grammar in Section~\ref{sec:udrive grammar}.
This is challenging to achieve with a language model alone, as they have the potential to hallucinate and make mistakes.
MLLMs such as ChatGPT-4, however, have added support for function calling~\cite{gpt_function_calling}, which enables users to connect the models with external tools and design integrated workflows. Additionally, OpenAI's introduction of `Structured Outputs' ensures that the arguments generated by the model adhere precisely to a specified JSON Schema, as defined by the user in the function call.

In \coolname, we implemented function calling based on a structured JSON Schema that describes the syntax of $\mu$Drive programs. This schema guides the MLLM to produce outputs that are always structured as syntactically valid $\mu$Drive programs, including all parameters and constructs mandated by the full $\mu$Drive grammar. By leveraging this function, we achieve a reliable and structured generation of $\mu$Drive scripts that align with expected syntax. For an example of function calling with our JSON Schema, please see our repository~\cite{source_code}.

Our function is designed with several key principles in mind:
1) \emph{Structural Integrity:} we ensure that the structure of the output $\mu$Drive program always adheres to the syntax of $\mu$Drive. Specifically, each program must include one trigger, zero or more conditions, one or more actions, and at most one exit trigger. The sequence is strictly enforced, i.e.~trigger, conditions, actions, followed by the exit trigger if it exists.
2) \emph{Comprehensive Descriptions:} to help the MLLM fully understand the meaning and functionality of each element, we add detailed descriptions to all events, conditions, and actions in natural language. These descriptions are sourced from the official documents of ADS and $\mu$Drive to ensure accuracy and clarity.
3) \emph{Clear Parameter Definitions:} the unit parameters within events, conditions, and actions are clearly defined to avoid any potential misunderstandings. This precision helps the MLLM to generate accurate and contextually appropriate programs.
By adhering to these design principles, integrating $\mu$Drive with an MLLM facilitates the creation of actionable and precise programs, that can then be applied in subsequent deployments of the ADS to improve its driving strategy.
For more information on our prompts, function calling, and implementation details, please refer to the 
source code~\cite{source_code}.

\section{Implementation and Evaluation}
\label{sec:implementation_evaluation}


\subsection{Implementation}
We have implemented \coolname for Apollo 9.0~\cite{apollo90} (the latest version at the time of our experiments) and the widely-used MLLM ChatGPT (version ChatGPT 4 Turbo). The simulator utilised in our experiments is the official \emph{Dreamview Plus~\cite{dreamview_plus}}, provided by Apollo 9.0. 

Our framework, \coolname, comprises three main components:
1) \emph{Trajectory Record Analysis Tool:} this tool identifies the specific time step at which the quantitative semantics of the trace fall below a specified threshold according to a given specification (such as no collisions and adherence to traffic rules in different countries). It enables us to precisely pinpoint the exact moment a violation occurs and to detect near-miss situations where a violation is likely but has not yet occurred.
2) \emph{Prompt Generator:} this component generates both the text prompt and captures a visualisation of the driving conditions at specific time steps. It organises this information into function calls for ChatGPT. Essentially, the the prompt generator automatically creates the input provided to the MLLM based on specified criteria and recorded data.
3) \emph{Translation and Verification:} the response from the MLLM is translated into a domain-specific language, $\mu$Drive, which outlines general driving strategy repairs (e.g.~stopping when a pedestrian is ahead). An additional validity check ensures that the syntax is correct.  These strategy repairs are then verified using the simulator to ensure they enable the autonomous vehicle to successfully navigate the given scenario.

Our implementation leverages some components from previous work. Specifically, from LawBreaker~\cite{Sun-Poskitt-et_al22a}, we utilise its specification language and the corresponding verification algorithm. From $\mu$Drive~\cite{wang2024mudrive}, we employ its DSL and backend support for applying new driving strategies in Apollo.

\subsection{Evaluation}
Our evaluation considers four Research Questions~(RQs):


\begin{itemize}
    \item \noindent \textbf{RQ1}: Does \coolname effectively repair the driving behaviour of an ADS?
    \item \noindent \textbf{RQ2}: Are these driving strategy repairs applicable to common driving scenarios?
    \item \noindent \textbf{RQ3}: How much effort is required to compute repairs?
    \item \noindent \textbf{RQ4}: What is the impact of using images for critical moments instead of text?
\end{itemize}
RQ1 considers whether \coolname achieves its primary goal of being able to utilise MLLMs to effectively repair the driving behaviour of an ADS following a violation event.
RQ2 investigates the effectiveness of \coolname's driving strategy repairs, based on a small sample of records, in improving AV performance across different scenarios. 
RQ3 examines the computational effort needed to utilise \coolname.
RQ4 validates the effects of using images in the prompt instead of ground-truth values in textual prompts.
Our experiments utilise both Apollo 9.0 and the Apollo Simulation Platform, referred to as Apollo Studio~\cite{Apollostudio}. To account for simulator randomness (e.g.~due to concurrency) each experiment is repeated 20 times, and we present the averages. 
We utilise a Linux machine (Ubuntu 20.04.5 LTS) with 32GB of RAM, an Intel i7-10700k CPU, and an RTX 2080Ti graphics card.

\begin{table*}
 \centering
 \caption{Performance comparison of \coolname, REDriver, and Apollo}
 \vspace{-0.2cm}\scriptsize
    \resizebox{0.9\linewidth}{!}{%
    \begin{tabular}{c|c|c|c|c|c|l}
    \hline
        Law & Scene & Driver & Fix & Pass & Robustness& \multicolumn{1}{c}{Context}\\
    \hline
    \multirow{6}{*}{no collision}  
    & \multirow{3}{*}{S1} & Apollo & -  & 0\% & \textcolor{red}{-0.1} &The AV entered the intersection during a green light vehicles, \\  
    & & REDriver & - & 0\% & \textcolor{red}{-0.1} & but failed to yield to the straight-moving\\  
    & & \coolname & 30\% & \textbf{100\%}  & \textcolor{green!50!black}{1.37}& resulting in an accident.\\
    \cline{2-7}
    & \multirow{3}{*}{S2} & Apollo & - & 0\% & \textcolor{red}{-0.1} &The AV fail to yield to the oncoming straight-through traffic\\  
    & & REDriver & - & 0\% & \textcolor{red}{-0.1} &at the stop sign and proceed to make a left turn at the intersection,\\  
    & & \coolname & 50\% & \textbf{100\%} & \textcolor{green!50!black}{4.41}&  resulting in an accident.\\
    \hline

    \multirow{6}{*}{\makecell{Law38}} 
    & \multirow{3}{*}{S3} & Apollo & - & 20\% & 10.7, \textcolor{red}{0.0}, \textcolor{red}{0.0}  &The AV started and entered the intersection when the traffic light\\
    & & REDriver & - & 60\% & 11.55, \textcolor{green!50!black}{0.5}, \textcolor{green!50!black}{0.5} & was yellow.\\ 
    & & \coolname & 45\% & \textbf{100\%} & 12.97, \textcolor{green!50!black}{0.5}, \textcolor{green!50!black}{0.48} &\\
    \cline{2-7}

    & \multirow{3}{*}{S4}  & Apollo & -  & 0\% & 4.7, 0.5, \textcolor{red}{0.0} &\multirow{3}{*}{The AV entered the intersection on a red light.}\\  
    & & REDriver & - & 85\% & 11.52, 0.5, \textcolor{green!50!black}{0.5} & \\  
    & & \coolname & 100\% & \textbf{100\%} & 2.84, 0.5, \textcolor{green!50!black}{0.5} &\\
    \hline
    
    \multirow{3}{*}{\makecell{Law44}}   
    & \multirow{3}{*}{S5} & Apollo & - & 20\% & \textcolor{red}{-19.98} &The AV is traveling in the fast lane and come to a stop due to\\  
    & & REDriver & - & \textbf{100\%} & \textcolor{green!50!black}{4.34} & an static obstacle (failure to change lanes to an available\\  
    & & \coolname & 20\% & \textbf{100\%} & \textcolor{green!50!black}{9.58}& lane on the right), ultimately failing to reach its destination.\\
    \hline

    \multirow{3}{*}{Law46}  & \multirow{3}{*}{S6} & Apollo & - & 0\%  &  \textcolor{red}{0.0}, \textcolor{red}{-0.2} & The AV continues to travel at speeds exceeding 30 kilometres \\  
    & & REDriver & - & \textbf{100\%} & \textcolor{green!50!black}{1.00}, \textcolor{green!50!black}{1.00} & per hour despite fog or rain.\\  
    & & \coolname & 100\% & \textbf{100\%}  & \textcolor{green!50!black}{1.23}, \textcolor{green!50!black}{1.23}&\\
    \hline

     \multirow{3}{*}{Law53}  & \multirow{3}{*}{S7} & Apollo & - & 0\%   &   \textcolor{red}{0.0}  &\multirow{3}{*}{The AV is approaching a junction with traffic jam.}\\  
     & & REDriver & - & 0\% & \textcolor{red}{0.0} & \\  
    & & \coolname & 50\% & \textbf{100\%} & \textcolor{green!50!black}{1.0}&\\
    \hline

    \multirow{3}{*}{finish journey}  & \multirow{3}{*}{S8} & Apollo & - & 0\%  &   \textcolor{red}{-0.42}  &The AV failed to overtake a stationary vehicle ahead and became\\  
    & & REDriver & - & 0\% & \textcolor{red}{-0.43} & stuck.\\  
    & & \coolname & 15\% & \textbf{100\%}  & \textcolor{green!50!black}{5.17}&\\
    \hline

    \end{tabular}}
    \label{tab:effectiveness comparison}
    \vspace{-0.5cm}
\end{table*}

\noindent \emph{\textbf {RQ1: Does \coolname effectively repair the driving behaviour of an ADS?}}
To answer this question, we employed a benchmark of scenarios provided by \cite{wang2024mudrive} where Apollo consistently violates specifications. 
Table~\ref{tab:effectiveness comparison} reports the effectiveness of our approach in preventing these violations compared to the original Apollo and the runtime enforcement method REDriver~\cite{sun2024redriver}. Note that the threshold for \coolname is set to 15, determined by an empirical experiment discussed later in this section.
The `Law' column in the table denotes the specific property specification under which the AV is tested. 
We adopted the formalisation of traffic laws reported in~\cite{Sun-Poskitt-et_al22a} as part of our property specifications and evaluated whether \coolname can be applied so that the ADS follows them. 
Specifically, we adopted four rules sourced from the \emph{Regulations for the Implementation of the Road Traffic Safety Law of the People's Republic of China}~\cite{China_traffic_law}: \emph{Law38}, \emph{Law44}, \emph{Law46}, and \emph{Law53}. These rules encompass regulations concerning traffic lights (yellow, green, red), speed limits for the fast lane, speed limits under adverse weather conditions (such as fog, rain, and snow), and managing traffic jam, respectively.
Additionally, we applied the property specifications `no collision' and `finish journey' to evaluate the AV as well. The detailed property specifications of these two rules can be expressed as: 
\vspace{-15pt}\begin{align*}
    no\_collision & \equiv \Box (\lnot NearestNPC(0.1)) \\
     finish\_journey & \equiv  \Box (\Diamond_{[0,200]}(speed > 0.5) \lor dest(5)) 
\end{align*}
Here, the specification `no collision' requires that the distance to other objects always be greater than 0.1 metres (i.e.~$\lnot NearestNPC(0.1)$). The specification `finish journey' requires that the AV must not stop on the road (i.e.~$\Diamond_{[0,200]}(speed > 0.5)$), unless it is close to the destination (i.e.~$dest(5)$). 
We refer the readers to our repository in \cite{source_code}
to see all the detailed specifications.
The `Fix' column in Table~\ref{tab:effectiveness comparison} shows the proportion of successful driving strategy repairs generated by \coolname. A driving strategy repair is deemed successful only if it ensures that the AV causes no violations.
To evaluate this, we repeat the generation process 20 times for each driving record, generating one unique driving suggestion per run. Each suggestion is then applied to the autonomous vehicle and tested for effectiveness in the same scenario through simulation, allowing us to assess whether the repair successfully resolves the failure.
Our empirical analysis indicates that while different suggestions may be generated for the same record, they typically converge into a limited set of outcomes. Consequently, 20 repetitions are empirically sufficient to capture all possible outcomes.
The `Pass' column indicates the proportion of runs that comply with traffic rules.
It signifies the success rate of each effective suggestion, which is always 100\% for \coolname.
Note that the MLLM generates varied suggestions across runs due to inherent uncertainty. 
However, empirical analysis shows that most effective suggestions are consistent across trials. Therefore, we select the most frequently successful suggestion to determine the final values in the `Pass' and `Robustness' column.
The `Robustness' column in the table illustrates the robustness of the AVs regarding current traffic regulations. The robustness value is calculated as the average performance of these effective suggestions. 
For example, the three values for the specification \emph{Law38} indicate the robustness values for green light, yellow light, and red light related traffic laws, respectively.
Specifically, the robustness value measures how closely the vehicle trajectory adheres to these rules. A higher robustness value indicates a lower likelihood of violating traffic regulations, whereas a lower value indicates a higher likelihood of imminent violation. A value less than or equal to $0$ indicates a violation of the corresponding traffic rule.
Furthermore, if a regulation comprises multiple sub-rules, the robustness for each sub-rule is sequentially presented.

\begin{figure}[t]
\begin{minipage}[t]{0.48\linewidth}
\begin{minted}[fontsize=\tiny,frame=single,obeytabs=true,tabsize=4,numbersep=5pt,escapeinside=||]{yaml}
|\textbf{\textcolor{green!50!black}{rule}}| |\textcolor{red!70!black}{"S1 rule1"}|
|\textbf{\textcolor{green!50!black}{trigger}}| 
    always 
|\textbf{\textcolor{green!50!black}{condition}}| 
    front_vehicle_closer_than(10)
|\textbf{\textcolor{green!50!black}{then}}|
    follow_dist(10)
    yield_dist(15)
    overtake_dist(20)
    obstacle_stop_dist(10)
    obstacle_decrease_ratio(1)
|\textbf{\textcolor{green!50!black}{end}}|
\end{minted}
\end{minipage}
\hfill
\begin{minipage}[t]{0.48\linewidth}
\begin{minted}[fontsize=\tiny,frame=single,obeytabs=true,tabsize=4,numbersep=5pt,escapeinside=||]{yaml}
|\textbf{\textcolor{green!50!black}{rule}}| |\textcolor{red!70!black}{"S1 rule2"}|
|\textbf{\textcolor{green!50!black}{trigger}}| 
    always 
|\textbf{\textcolor{green!50!black}{condition}}| 
    is_traffic_light(red)
    traffic_light_distance_leq(10)
|\textbf{\textcolor{green!50!black}{then}}|
    traffic_light_stop_dist(5)
|\textbf{\textcolor{green!50!black}{end}}|
\end{minted}
\end{minipage}
\caption{\textsc{$\mu$Drive} driving strategy repair scripts for S1}
\label{lst:s1 commands}
\vspace{-0.6cm}
\end{figure}

As shown in Table~\ref{tab:effectiveness comparison}, Apollo's success rate in these scenarios consistently remains below 50\%, often reaching 0\%. While REDriver can prevent some of the violations sometimes, there are still instances it cannot address.
In contrast, the repairs by \coolname enable the AV to completely avoid accidents and violations.
This is because REDriver focuses on a narrow case-by-case view, making decisions based only on current perception and prediction, and thus heavily depends on the accuracy of the ADS's original predictions, which \emph{may be wrong}.
For example, in scenario \emph{S1} where an AV fails to yield to a straight-moving vehicle, REDriver cannot prevent the violation because it cannot reverse the decision `not to yield', which was based on the prediction that the vehicle would not obstruct its path. 
In contrast, in the driving strategy repairs generated by \coolname, as shown in Figure~\ref{lst:s1 commands}, the first program dynamically adjusts parameters such as follow distance, yield distance, overtake distance, stop distance, and obstacle response rate when a vehicle is within 10 metres ahead. Additionally, the second program ensures safety by adjusting the AV's stop distance when approaching a red light and the distance to the stop line is less than 10 metres. This adaptive approach, based on a global perspective, continuously modifies the vehicle's driving style as conditions change, thereby effectively avoiding accidents before they occur. 
Due to the inherent uncertainty of the generative model, the proportion of successful repairs sometimes falls below $50\%$ (as shown in `Fix' column).  However, our evaluation of \coolname's time and cost efficiency in RQ3 demonstrates that generating each driving strategy repair is both time-efficient and cost-effective, mitigating this issue.

\begin{table}[t]
\centering
\caption{Effectiveness of \coolname across varying $\delta$}
\vspace{-0.2cm}\scriptsize
\label{tab:thresholds}
\resizebox{0.7\linewidth}{!}{
\begin{tabular}{|c|cccccccc|}
\hline
$\delta$ & S1 & S2 & S3 & S4 & S5 & S6 & S7 & S8 \\
\hline
 1 & $\times$ & $\times$ & $\times$ & \checkmark & \checkmark & \checkmark & \checkmark & $\times$ \\ \hline
 5 & $\times$ & $\times$ & \checkmark & \checkmark & \checkmark & \checkmark & \checkmark & $\times$ \\ \hline
 10 & $\times$ & \checkmark & \checkmark & \checkmark & \checkmark & \checkmark & \checkmark & \checkmark \\ \hline
 15 & \checkmark & \checkmark & \checkmark & \checkmark & \checkmark & \checkmark & \checkmark & \checkmark \\ \hline
 20 & $\times$ & \checkmark & \checkmark & \checkmark & \checkmark & \checkmark & \checkmark & $\times$ \\ \hline
 25 & $\times$ & \checkmark & \checkmark & \checkmark & \checkmark & \checkmark & \checkmark & $\times$ \\ \hline
 30 & $\times$ & \checkmark & \checkmark & \checkmark & \checkmark & \checkmark & \checkmark & $\times$ \\
\hline
\end{tabular}
}
\end{table}

To further investigate RQ1, we designed a second experiment to examine how varying \coolname's thresholds impact its effectiveness. Recall that the threshold defines the fault tolerance of \coolname, determining what constitutes a near-miss situation, as discussed in Section~\ref{sec:Violation and Near-Miss Moment}. The detailed effectiveness evaluation for all thresholds is shown in Table~\ref{tab:thresholds}. In this table, the first column (`$\delta$') denotes the threshold value, ranging from 1 to 30. The subsequent columns represent scenarios S1 to S8 as mentioned above. If \coolname can provide driving strategy repairs that help the AV resolve the encountered problem, i.e.~satisfy the corresponding specification within 10 queries, we mark it with a \checkmark. Otherwise, we mark it with a $\times$. 
As show in Table~\ref{tab:thresholds}, the threshold value significantly impacts \coolname's effectiveness in certain scenarios. When the threshold $\delta$ is very small, such as 1 or 5, the near-miss moments are too close to the violation moment, providing insufficient information about why the violation occurs. Conversely, if $\delta$ is too large, such as 30, the near-miss moment may not provide any useful information, as there may be no indication of a potential violation. This can lead to \coolname's failure to deliver effective programs.
For example, in scenario S1, where the AV is making a right turn and fails to yield to vehicles going straight, the timing of the threshold is critical. When the threshold is set to values below 10, the AV is already impeding the straight-moving vehicles at the near-miss moment, making it difficult to implement effective changes. Conversely, when the threshold is set to 20, the AV has not started the right turn yet at the near-miss moment, and the potential problem has not yet arisen. In some scenarios, \coolname can be effective for all threshold values if the near-miss moment does not provide critical information. For example, in scenario S6, where the vehicle exceeds 30 km/h in snowy conditions, the critical information is that the AV exceeds 30 km/h at the violation moment. Since snow conditions remain consistent, the choice of near-miss moment does not affect \coolname's effectiveness.
\begin{table}[]
    \centering
    \caption{Performance of \coolname in official scenarios}
    \vspace{-0.2cm}\scriptsize
    \begin{tabular}{c|c|c|c|c}
    \hline
         \multirow{2}{*}{Map} & \multirow{2}{*}{Num} & \multirow{2}{*}{Driver} & \multirow{2}{*}{Finish} & \multirow{2}{*}{Accident} \\
         & & & & \\
         \hline
         \multirow{2}{*}{Sunnyvale} & \multirow{2}{*}{114} & Apollo & 108 & 7   \\
         \cline{3-5}
         & & \coolname & 112 & 6  \\
         \hline
         \multirow{2}{*}{SanMateo} & \multirow{2}{*}{103} & Apollo & 94 & 1 \\
         \cline{3-5}
         & & \coolname  & 95  & 1 \\
         \hline
         \multirow{2}{*}{Apollo Virtual} & \multirow{2}{*}{52} & Apollo & 42 & 1  \\
         \cline{3-5}
         & & \coolname & 46 & 1 \\
         \hline
    \end{tabular}
    \vspace{-0.5cm}
    \label{tab:comparison_with_official_scenarios1}
\end{table}

\begin{table*}[]
    \centering
    \caption{Performance comparison of \coolname and Apollo in official scenarios}
    \vspace{-0.2cm}\scriptsize
    \begin{tabular}{c|c|c|c|c|c|c|c|c|c|c}
    \hline
         \multirow{2}{*}{Map} & \multirow{2}{*}{Num} & \multirow{2}{*}{Driver} & \multicolumn{2}{c|}{Speed($m/s$)} & \multicolumn{2}{c|}{Accelerate($m/s^2$)} & \multicolumn{2}{c|}{Obstacle Distance($m$)} & \multirow{2}{*}{Stop Time($s$)} & \multirow{2}{*}{Energy($J$)}\\
         \cline{4-9}
         & & &  Ave & Max & Ave & Max & Ave & Min & &\\
         \hline
         \multirow{2}{*}{Sunnyvale} & \multirow{2}{*}{108} & Apollo & 3.31 & 7.86 & 0.38 & 1.91 & 51.43 & 6.74 & 15.56 & 132840.52  \\
         \cline{3-11}
         & & \coolname & 3.76 & 8.29 & 0.46 & 2.19 & 51.71 &	6.90 & 12.63 & 147434.25 \\
         \hline
         \multirow{2}{*}{SanMateo} & \multirow{2}{*}{94} & Apollo & 2.58 & 6.28 & 0.41 & 1.47 & 31.27 & 6.85 & 6.95 & 70570.81 \\
         \cline{3-11}
         & & \coolname  & 2.84 &	6.20 & 0.44 & 1.80 & 30.97 & 7.45 & 4.48 & 69660.75 \\
         \hline
         \multirow{2}{*}{Apollo Virtual} & \multirow{2}{*}{42} & Apollo & 4.10 & 8.43 & 0.43 & 1.75 & 78.48 & 4.06 & 13.95 & 157633.50 \\
         \cline{3-11}
         & & \coolname & 4.59 & 9.18 & 0.50 & 1.95 & 78.22 & 4.39 & 6.00  & 210024.93\\
         \hline
    \end{tabular}
    \vspace{-0.5cm}
    \label{tab:comparison_with_official_scenarios2}
\end{table*}

\noindent \emph{\textbf {RQ2: Are these driving strategy repairs applicable to common driving scenarios?}}
To answer this question, we applied all the $\mu$Drive driving repair scripts generated by \coolname across the eight scenarios mentioned above. In total, there are 22 different driving strategy repair programs. 
For detailed information on these driving strategy repair programs, we refer readers to \cite{source_code}.
We applied these strategy repairs to all the official scenarios provided by Apollo across three maps: \emph{Sunnyvale}, \emph{San Mateo}, and \emph{Apollo Virtual}.
For the \emph{Sunnyvale} map, there are 114 different scenarios. The \emph{San Mateo} map contains 103 different scenarios, while the \emph{Apollo Virtual} map includes 52 different scenarios. These scenarios cover most situations encountered during daily city driving, such as passing traffic lights, yielding to pedestrians and priority vehicles, cutting in, changing lanes, overtaking, and making U-turns. For detailed descriptions of these official scenarios provided by Apollo, refer to~\cite{Apollostudio}. 

First, we compared Apollo with \coolname (i.e.~Apollo with the driving repairs applied) regarding \emph{Completion Rate} and \emph{Accidents}, as shown in Table~\ref{tab:comparison_with_official_scenarios1}.  Regarding \emph{Completion Rate}, we evaluated whether the AV successfully reached the destination and completed the journey, as indicated in the `Finish' column. All scenarios completed by Apollo were also completed by \coolname. Additionally, \coolname could finish extra scenarios where Apollo got stuck. For example, in some scenarios, Apollo failed to overtake a stationary vehicle ahead because it followed too closely, while \coolname completed these scenarios by maintaining a larger following distance. Regarding \emph{Accidents}, we examined the number of accidents caused by Apollo and \coolname, as shown in the `Accident' column. \coolname not only avoided causing new accidents but also prevented an accident in one scenario. It is important to note that all remaining accidents were not caused by the AV. They were caused by `irrational' vehicles or pedestrians colliding with the driver from behind. Typically, this occurred when the AV had reached its destination and stopped, and another vehicle hit it from the back.

To further investigate RQ2, we conducted an in-depth analysis. Specifically, for scenarios that Apollo and \coolname successfully completed, we analysed various aspects of the trajectories, including the speed and acceleration at each point, the distance to the nearest obstacle, the total duration of vehicle stops, and the energy consumption, as shown in Table~\ref{tab:comparison_with_official_scenarios2}.
Regarding \emph{Speed and Acceleration}, we compared the speed and acceleration between Apollo and \coolname, as shown in the `Speed($m/s$)' and `Acceleration($m/s^2$)' columns. Both the average and maximum speeds were examined, with the values in the table representing the averages across all scenarios. Overall, the AV operated at slower speeds under Apollo compared to \coolname. However, this does not imply that \coolname drives more aggressively than Apollo. In fact, \coolname only increased its speed when there were no other vehicles or pedestrians nearby, ensuring safe and considerate driving behaviour.
Regarding \emph{Obstacle Distance}, we examined the average and maximum distances to other objects, as shown in the `Obstacle Distance(m)' column. The results indicated that \coolname maintained a greater distance from other vehicles despite its higher speed, demonstrating both efficiency and safety.
Regarding \emph{Stop Time}, we examined the time that the AV stopped on the road, as shown in the `Stop Time(s)' column. The results indicated that \coolname had less stop time than Apollo, suggesting a smoother driving experience.
Regarding \emph{Energy}, we provided a rough estimate of the average energy consumption for Apollo and \coolname. The energy was calculated using the formula: $\sum_{t=1}^{n-1} \frac{1}{2}m (v_{t+1}^2 - v_t^2)$, where $m$ is the vehicle's mass (1500 kg), $n$ is the length of the trace, and $v_t$ is the speed of the AV at time step $t$. This formula measures the energy consumption based on changes in speed. \coolname consumed more energy than Apollo because it generally travelled at higher speeds, which involved more frequent acceleration and deceleration processes.

Based on these detailed checks, we conclude that the driving strategy repairs provided by \coolname not only promote smoother driving but also contribute to fewer accidents, underscoring its suitability for various common driving scenarios.

\begin{table}[ht]
\centering
\caption{Computational effort required by \coolname}
\vspace{-0.2cm}\scriptsize
\label{tab:effort}
\begin{tabular}{|>{\centering}m{0.05\linewidth}|>{\centering}m{0.058\linewidth}|>{\centering}m{0.05\linewidth}|>{\centering}m{0.05\linewidth}|>{\centering}m{0.05\linewidth}|>{\centering}m{0.05\linewidth}|>{\centering}m{0.05\linewidth}|>{\centering}m{0.05\linewidth}|>{\centering}m{0.05\linewidth}|m{0.05\linewidth}|}
\hline
\multicolumn{2}{|c|}{step} & S1 & S2 & S3 & S4 & S5 & S6 & S7 & S8 \\
\hline
 \multicolumn{2}{|c|}{trace} & 329s & 351s & 281s & 127s & 61s & 63s & 395s & 529s \\ \hline
 \multicolumn{2}{|c|}{localisation} & 187s & 200s & 167s & 156s & 134s & 155s & 205 & 168 \\ \hline
 \multicolumn{2}{|c|}{prompt} & 0.22s & 0.24s & 0.21s & 0.23s & 0.16s & 0.17s & 0.16s & 0.22 \\ \hline
 \multirow{3}{*}{query} & time & 10.6s & 11.4s & 14.1s & 8.9s & 14.9s & 8.6s & 23.3s & 10.8s \\ \cline{2-10}
 & input & 7352 & 7352 & 7436 & 7435 & 7508 & 7498 & 7504 & 7350 \\ \cline{2-10}
 & output & 179 & 163 & 121  & 185 & 97 & 81 & 123 & 82 \\ \hline
  \multicolumn{2}{|c|}{overall time} & 527s & 563s & 462s & 292s & 210s & 227s & 624s & 708s \\ \hline
  \multicolumn{2}{|c|}{cost(\$)} & 0.079 & 0.078 & 0.078 & 0.080 & 0.078 & 0.077 & 0.079 & 0.076 \\ \hline
\end{tabular}
\vspace{-0.1cm}
\end{table}

\noindent \emph{\textbf {RQ3: How much effort is required to compute repairs?}} To answer this question, we present a detailed breakdown of time and token consumption (using model ChatGPT 4 turbo) for \coolname, as shown in Table~\ref{tab:effort}.
The `step' column lists all necessary steps for \coolname to generate driving strategy repairs. The `trace' step involves converting a given record into a trace. The `localisation' refers to identifying \emph{near-miss} and \emph{violation} moments. The `prompt' involves automatically generating prompts for the LLM input, while `query' denotes querying the LLM for a response. 

We detail the time consumption for each step, all measured in seconds. For the whole process, the most time-consuming steps involve two parts: trace generation and moment localisation. Trace generation takes a few minutes due to the thousands of time steps within a trace, typically about one hundred time steps per second. Moment localisation involves calculating the robustness value multiple times, resulting in relatively high time consumption. However, since our method is offline, trace generation and moment localisation need to be performed only once per test case, making the process efficient for practical use. For each test case, the whole process typically takes around 10 minutes to perform, always below 15 minutes, on a desktop with 32GB of RAM, an Intel i7-10700k CPU, and an RTX 2080Ti graphics card, which is a manageable effort.

Additionally, we measure the number of tokens required for querying the LLM. The `input' and `output' rows in the Table~\ref{tab:effort} indicate the average number of input and output response tokens for ChatGPT 4 turbo.  
The number of tokens, including those for images, is calculated using ChatGPT's official tool~\cite{tokenizer_web}.
At the time of experimentation, the direct cost for 1 million input prompt tokens was \$10, while 1 million output response tokens cost \$30. This indicates that each driving suggestion costs less than \$0.08, making it affordable, as shown in the last row of the table.

\begin{table}[h]
\centering
\caption{Comparison of \coolname and a text-only method}
\vspace{-0.2cm}\scriptsize
\label{tab:images_text_transposed}
\begin{tabular}{|c|c|c|c|c|c|c|c|c|}
\hline
\multirow{2}{*}{Scene} & \multicolumn{2}{c|}{performance} & \multicolumn{2}{c|}{input token} & \multicolumn{2}{c|}{output token} & \multicolumn{2}{c|}{cost(\$)} \\ \cline{2-9}
 & ours & text & ours & text & ours & text & ours & text\\ 
\hline
S1 & $30\%$ & $5\%$     & 7352  & 8370 & 179& 308 & 0.079 & 0.092\\ \hline
S2 & $50\%$ & $0\%$     & 7352  & 7294 & 163& 286 & 0.078 & 0.082\\ \hline
S3 & $45\%$ & $5\%$     & 7436  & 8451 & 121& 245 & 0.078 & 0.092\\ \hline
S4 & $100\%$ & $15\%$   & 7435  & 7741 & 185& 247 & 0.080 & 0.085\\ \hline
S5 & $20\%$ & $0\%$     & 7508  & 8442 & 97 & 294 & 0.078 & 0.093\\ \hline
S6 & $100\%$ & $100\%$  & 7498  & 7433 & 81 & 189 & 0.077 & 0.080\\ \hline
S7 & $50\%$ & $20\%$    & 7504  & 10978& 123& 351 & 0.079 & 0.120\\ \hline
S8 & $15\%$ & $0\%$     & 7350  & 7240 & 82 & 241 & 0.076 & 0.080\\ \hline
\end{tabular}%
\vspace{-0.3cm}
\end{table}

\noindent \emph{\textbf {RQ4: What is the impact of using images for critical moments instead of text?}}
\coolname utilises visualisations of violation and near-miss moments as part of its prompt for the MLLM. 
But what would happen if we described these scenarios using only textual prompts instead? To explore this question, we establish a text-only prompt-based method as our baseline. 
To ensure a fair comparison, we keep all other design elements consistent with \coolname, except that descriptions of the violation and near-miss moments are provided solely in text. 
We extract key information from records following the LawBreaker methodology~\cite{Sun-Poskitt-et_al22a}, crafting detailed descriptions for each variable to ensure clarity. These descriptions are formatted and refined using ChatGPT-4 Turbo, optimising them for MLLM interpretation. An example prompt is available in our repository for reference~\cite{source_code}.

Table~\ref{tab:images_text_transposed} compares \coolname and the text-only method in terms of performance, input/output token usage, and cost. The performance threshold, set at 15 based on empirical experimentation (discussed in RQ1), includes 20 repetitions per scenario. The `Performance' column shows the proportion of successful driving strategy repairs generated by \coolname (referred to as `ours') and the text-only method (referred to as `text'). Here, success indicates that the AV correctly follows the traffic rule after the repair. The `Input/Output Tokens' columns display the average input and output token counts per query. As shown, using images significantly enhances performance while reducing costs per query. Images effectively convey spatial details that are challenging to capture in text yet are easily processed by the vision modality. 
Interestingly, image prompts consume fewer input tokens than detailed text descriptions. 
Moreover, text-heavy prompts often result in more output tokens, suggesting that the MLLM is more prone to generating extraneous driving strategy repairs when overloaded with extensive text inputs.

\noindent \emph{\textbf {Threats to Validity. }}
The inherent randomness of generative models and the limitations of the original ADS introduce threats to validity. First, \coolname cannot guarantee the effectiveness of every generated suggestion. To mitigate this, we generate 20 driving strategy repairs per case and evaluate them in an AV simulator, leveraging fast and cost-effective querying for robustness.

Some scenarios remain beyond \coolname's full control, such as rear-end collisions, where following distance depends on the trailing vehicle. While risk-reduction measures exist, complete prevention is challenging. Moreover, \coolname may propose valid repairs that prove ineffective due to ADS design constraints. For instance, Apollo’s overly cautious behaviour might prevent overtaking, even when \coolname suggests it. Refining text prompts can help address such issues.

Integrating $\mu$Drive into an ADS poses challenges, but once incorporated, it streamlines further modifications, enabling efficient system refinement through various $\mu$Drive scripts.

\section{Related Work} \label{sec:related}
AVs have been the subject of extensive research in recent years, leading to significant advancements in their capabilities. Early efforts in AV development focused on improving core functionalities such as perception, planning, and control~\cite{levinson2011towards, yurtsever2020survey}. These areas are critical for enabling AVs to navigate complex environments safely. However, the limitations of AVs compared to human drivers, particularly regarding adaptability and decision-making in unpredictable scenarios, have prompted further research into more intelligent systems.

Several approaches have been proposed to address challenges encountered by AVs at runtime. Rule-based systems have been used to ensure adherence to safety and traffic regulations. For example, runtime enforcement mechanisms monitor the vehicle's actions~\cite{Mauritz-et_al16a, d2005lola, Watanabe-et_al18a} to prevent collisions and other unsafe behaviours~\cite{Grieser-et_al20a, hong2020avguardian, Cheng-et_al21a, Shankar-et_al20a}. Similarly, gradient-based algorithms, such as those in REDriver~\cite{sun2024redriver}, offer real-time solutions for handling property specification violations. While these methods provide valuable safety nets, their utility is limited by their narrower focus on specific tasks.

Recognising the expertise of human drivers, researchers have explored various ways to model and replicate human driving behaviour in AVs. Imitation learning has emerged as a prominent technique for training AVs to mimic expert human drivers~\cite{sama2020extracting, wei2010learning, xu2020learning}. These methods aim to capture the nuanced decision-making processes of human drivers to improve AV performance. However, challenges such as limited training data and the complexity of human driving behaviour have hindered the generalisation of these approaches~\cite{le2022survey}. As a result, there is a growing interest in developing more advanced systems that can better bridge the gap between human intelligence and AV technology.

The advent of MLLMs has opened new avenues for enhancing AV intelligence. MLLMs, with their advanced text and image understanding capabilities, offer promising solutions for interpreting and replicating human driving behaviour. They can provide natural language explanations for their decisions, thereby enhancing transparency and trust~\cite{cui2024survey}. Existing research has explored the use of MLLMs in various components of AVs, including perception, planning, and control~\cite{chen2023driving, mao2023gpt, wen2024road}. For instance, LLM-Driver~\cite{chen2023driving} abstracts driving scenarios into 2D object-level vectors and directly applies the LLM output as control commands for the AV system. GPT-Driver~\cite{mao2023gpt} translates motion planner inputs and outputs into language tokens, utilising LLMs as motion planners for AVs. Wen et al.~\cite{wen2024road} evaluated the potential of ChatGPT-4 as an autonomous driving agent, demonstrating its advanced scene understanding and causal reasoning capabilities. 
These works primarily employ LLMs for object perception, motion planning, and actuation control within AV systems. Despite their potential, the inherent delays and uncertainties associated with generative models pose challenges for real-time AV operations. Additionally, the gap between natural language and control commands remains a significant hurdle.

\coolname, in contrast, is a framework that generates driving strategy repairs for AVs. By providing a general offline solution, it ensures that MLLMs can generate general driving suggestions that are directly applicable to AVs. Through comprehensive testing in various scenarios, \coolname has demonstrated its effectiveness in improving AV decision-making and adherence to property specifications, offering an advancement to the field of autonomous driving.

\section{Conclusion}
\label{sec:conclusion}
We have proposed \coolname, a framework that uses MLLMs to enhance ADSs by generating intelligent driving strategy repairs. \coolname identifies critical moments in driving scenarios and generates prompts to ensure MLLMs produce valid suggestions in a DSL, $\mu$Drive, for direct application in an ADS.
Experimental results show that \coolname improves ADS performance in various challenging scenarios, providing efficient and cost-effective driving suggestions. 
This framework represents a step towards bridging the gap between human expertise and automated driving technology, enhancing the adaptability and reliability of AVs. 

\section*{Acknowledgment}
This research is supported by the Ministry of Education, Singapore under its Academic Research Fund Tier 3 (Award ID: MOET32020-0004). Any opinions, findings and conclusions or recommendations expressed in this material are those of the author(s) and do not reflect the views of the Ministry of Education, Singapore.

\bibliographystyle{IEEEtran}
\bibliography{reference}

\end{document}